%% file: main.tex
\documentclass[
twocolumn,
]{ceurart}

\usepackage{xurl}

\begin{document}
\sloppy

\pretolerance=6000
\tolerance=2000 
\emergencystretch=10pt 
\copyrightyear{2021}
\copyrightclause{Copyright for this paper by its authors.
  Use permitted under Creative Commons License Attribution 4.0
  International (CC BY 4.0).}

\conference{Joint Proceedings of the ACM IUI 2021 Workshops,
  April 13--17, 2021, College Station, USA}

\title{Retrofitting Meetings for Psychological Safety}

\author[1]{Marios Constantinides}[%
orcid=0000-0003-1454-0641,
]
\ead{marios.constantinides@nokia-bell-labs.com}

\author[1]{Sagar Joglekar}[%
orcid=0000-0002-8388-9137,
]
\ead{sagar.joglekar@nokia-bell-labs.com}

\author[1, 2]{Daniele Quercia}[%
orcid=0000-0001-9461-5804,
]
\ead{daniele.quercia@nokia-bell-labs.com}
\address[1]{Nokia Bell Labs, Cambridge, UK}
\address[2]{King’s College, London, UK}


\begin{abstract}
Meetings are the fuel of organizations' productivity. At times, however, they are perceived as wasteful vaccums that deplete employee morale and productivity. Current meeting tools, to a great extent, have simplified and augmented the ways meetings are conducted by enabling participants to ``get things done'' and experience a comfortable physical environment. However, an important yet less explored element of these tools' design space is that of psychological safety---the extent to which participants feel listened to, or motivated to be part of a meeting. We argue that an interdisciplinary approach would benefit the creation of new tools designed for retrofitting meetings for psychological safety. This approach comes with not only research opportunities---ranging from sensing to modeling to user interface design---but also challenges---ranging from privacy to workplace surveillance.
\end{abstract}


\maketitle

\input{sections/1_Introduction}
\input{sections/2_HumanComms}



\bibliography{main}


\end{document}

%% file: sections/1_Introduction.tex
\section{Introduction}
\label{sec:introduction}
Meetings are often considered as the fuel of an organization's productivity. Employees come together for a common purpose to discuss ideas, to make collective decisions, and to ultimately reach their objectives. However, it is no secret that meetings are often seen as wasteful vacuums, or as an enemy of productivity. Although there are good meetings and bad meetings, their collective negative impact on employee morale and productivity is significant~\cite{kauffeld2012meetings}. To moderate this, organizations devote notably large amounts of resources to facilitate and support them. While meeting tools pledge to facilitate better meetings---ones that are well-executed, and create a safe environment for contribution---yet report after report estimate a growth of ineffective meetings\footnote{\url{https://www.forbes.com/sites/danielreed/2019/01/30/report-suggests-that-23rds-of-the-100-billion-spent-annually-on-business-meetings-travel-is-wasted}}; numbers though that are bound to change, if meeting tools were to fully capture people's meetings experience. Recently, in a large-scale crowdsourcing study~\cite{comfeel}, researchers determined the main factors that impact people's meetings experience. They administered a 28-item questionnaire to 363 individuals whose answers were statistically analyzed through Principal Component Analysis, and found that three factors sufficiently capture people's experience in meetings, namely, (a) execution, (b) physical comfort, and (c) psychological safety. Put differently, these factors capture the extent to which (a) people feel that a meeting was productive, (b) the meeting room was pleasant, and (c) the setting was psychologically safe.

Current meeting tools, however, primarily focus on enabling participants to ``get things done'' (i.e., \emph{execution}), or on optimizing the environmental conditions (i.e., \emph{physical comfort}). \textbf{Execution} is about whether a meeting had a clear structure, purpose, and resulted into a list of actionable items; a large body of previous research focused on these aspects. Kim and Rudin~\cite{kim2014learning} developed a system that detects key decisions in dialogues, while McGregor and Tang~\cite{mcgregor2017more}'s system generates an `action items' list from the spoken dialogue. Using an agenda planning technique, Garcia et al.~\cite{garcia2004cutting} allowed meetings participants to vote for agendas to improve perceived meeting quality. Video Threads~\cite{barksdale2012video} provides asynchronous video sharing, while Time Travel Proxy~\cite{tang2012time} identifies the gist of what was missed to enable late participation effectively. \textbf{Physical comfort} is about whether the meeting room was pleasant (in regard to air quality and crowdedness). In Human-Building Interaction literature~\cite{alavi2019introduction}, poor environmental conditions are known to impact employees' cognitive functions, decision making, and performance. Therefore, organizations have resorted to sensors through which the environmental conditions could be sensed~\cite{comfeel, alavi2017comfort}, and even adapted accordingly~\cite{bader2019windowwall} to meet recommended standards, thus increasing their employees’ productivity and well-being. To this end, meeting tools could account for physical comfort in their design, potentially in a form of interventions (e.g., adjusting ventilation). However, \emph{the design space should not only facilitate execution and ensure that desirable physical environmental conditions are met, but should also cultivate a psychologically safe setting.}

%% file: sections/2_HumanComms.tex
\section{Capturing psychological safety}
\label{sec:human_commms}

\textbf{Psychological safety} is about whether participants felt listened to, or motivated to be part of a meeting. Edmondson described it ``as the absence of interpersonal fear that allows people to speak up with work-relevant content''~\cite{edmondson1999psychological}. Previous research showed that inclusiveness and balance of conversational turn-taking play an important role in group performance~\cite{woolley2010evidence}. Tools have been developed to create awareness by highlighting salient moments during conversations~\cite{bergstrom2009conversation}, to enhance group collaboration through persuasive feedback~\cite{kim2008meeting}, and to allow participants to reflect on their own and their peers' experiences~\cite{meetcues}. Although such tools, to some extent, offer features that facilitate execution and support physical comfort, they often fall short in enabling psychological safety. We argue that a more interdisciplinary approach would likely benefit the creation of new tools designed for supporting psychological safety. To achieve that, we foresee a number of challenges and opportunities, ranging from sensing to modeling to user interface (UI) design.

\textbf{Sensing.} New emerging sensing devices such as earables~\cite{kawsar2018earables} are now fully equipped with IMU (inertial measurement unit) sensors, allowing on-body and on-device sensing. This opens up a new avenue for meeting tools by allowing participants to monitor signals that could otherwise go unnoticed; for example, capturing body cues of (dis)agreement or (in)active participation during a virtual conversation when the video stream is absent~\cite{kairos}. Similarly, smartwatches are now fully equipped with heart rate sensors that provide a window to people's physiology, allowing one to track their own or their peers' emotional states~\cite{park2020wellbeat}; aspects that are closely linked to creating a safe environment for contribution. Additionally, in the future, we foresee that better precision devices would enable more nuanced non-verbal or verbal communication patterns to be captured.

\textbf{Modeling.} New algorithms are also likely to provide a new understanding and perspective that would help further theorize the concept of psychological safety. New Natural Language Processing text-mining algorithms~\cite{choi2020ten} are now able to reveal certain language markers that might be deeply hidden in a conversation, particularly in a remote setting. These algorithms can annotate everyday language and capture important types of social interactions (e.g., a heated discussion resulting in conflict resolution). NLP-based algorithms can now analyze conversations and test whether these conversations accommodate different points of view~\cite{robertson2019language}, or even reveal the presence (or absence) of certain health-related markers (e.g., stress markers)~\cite{scepanovic2020extracting}. Additionally, new Natural Sound Processing algorithms~\cite{curtis2015effects} are now able to model verbal cues (e.g., prosody) that would potentially enable richer and more focused interactions. For example, prosodic features (e.g., pitch and energy) are known to provide a reliable indication of the emotional status in a conversational exchange.

\textbf{UI design.} Last but not least, new opportunities are likely to arise for the UI design community. New visualizations are more likely to be (re)invented, beyond dashboards and simple analytics~\cite{few2006information}. Drawing from behavioral economics research, we foresee that new forms of interventions would allow people to be more empathetic, compassionate, and aware of each other's emotional states, views, and thoughts. Previous research in the area of organizational behavior showed that affective sharing within groups conveys our internal experiences, signals our emotional states and, potentially, makes us more aware and empathic of each other~\cite{walter2008positive}. Borrowing ideas from calm technology~\cite{weiser1997coming} and biophilic design, meeting tools could embrace new types of cues only available through technology~\cite{qin2020heartbees}. For example, the use of different symbols, imagery, and artificial artifacts (e.g., real-world objects~\cite{yu2017stresstree}, light~\cite{yu2018delight},
or movement~\footnote{Blooming: \url{http://www.thelisapark.com/blooming}}) could augment the ways we interact and communicate with each other. These new visualizations could bring teams together despite working apart, and remove any geographical barriers due to physical distancing.

\textbf{Workplace surveillance.} While this interdisciplinary approach promises to deliver experiences richer of psychological safety, it also raises questions relating to workplace surveillance. It is often regarded that organizations and surveillance go hand in hand~\cite{ball2010workplace}. On a very pragmatic level, there is a handful of reasons as to why organizations opt in for employees' surveillance (e.g. maintaining productivity, monitoring resources used, protecting the organization from legal liabilities). The critics, however, rightly argue that there is a fine line between what organizations could be monitoring and what they should be monitoring. If crossed, it will have consequences on employees, affecting their well-being, work culture, and productivity. If future meetings tools incorporate any kind of employees' monitoring, they need to ensure that is done in a way that preserves an individual's rights, including that of privacy.

The workplace is constantly changing and evolving. These changes are currently accelerated by the COVID-19 pandemic, which might be leading to a dramatic change in how we work: from the well-known eight-hour workday to the office building to the salient boundaries between work and personal life. Meetings are no exception to this sudden change. In a post-pandemic world, we envision that new sensing devices would provide access to employees' data that otherwise might not be possible to collect (e.g., on-body sensing); that new algorithms would `make sense' of such data, and capture behavioral aspects that are hard to quantify (e.g., (dis)agreement, empathy, or stress markers); and that new user interfaces (e.g., inspired by biophilic design) would enable meeting participants to stay connected despite any geographical or technological barriers due to remote working.